\documentclass[onecolumn,authoryear]{els-mrw}

\usepackage{graphicx} 

\usepackage{url}
\usepackage{comment}
\usepackage{color}

\usepackage{hyperref}
\hypersetup{
    colorlinks = true,
    citecolor  = blue,
    linkcolor  = blue,
    urlcolor  = blue
}

\usepackage{mathtools, braket}
\usepackage{amsmath,amsfonts,amssymb,amsthm}

\usepackage[most]{tcolorbox}
\usepackage{xcolor}

\newtcolorbox{keypoint}{
  colback=gray!15,   
  colframe=gray!60,  
  boxrule=0pt,       
  left=6pt,
  right=6pt,
  top=6pt,
  bottom=6pt
}

\title{EncyclopediaNP}

\begin{document}
\chapter{Parton distribution functions from lattice QCD}
\author[1]{Martha Constantinou\footnote{Principal author of this review. Contact information: marthac@temple.edu}}
\author[2]{Krzysztof Cichy}

\address[1]{\orgname{Temple University}, \orgdiv{Department of Physics}, \orgaddress{Philadelphia, PA 19122 - 1801, USA}}
\address[2]{\orgname{Adam Mickiewicz University}, \orgdiv{Faculty of Physics and Astronomy}, \orgaddress{ul.\ Uniwersytetu Poznańskiego 2, 61-614 Poznań, Poland} }

\maketitle

\begin{abstract}
Parton distribution functions (PDFs) provide one of the most direct ways to describe the partonic structure of hadrons in QCD. They encode nonperturbative information about quarks, antiquarks, and gluons as functions of the partonic momentum fraction $x$, and they connect this microscopic structure to experimentally measurable high-energy scattering processes through QCD factorization. This makes PDFs interesting to pursue with lattice QCD, which provides a first-principles formulation of the strong interaction. The challenge is that PDFs are defined through light-cone correlations, while lattice QCD is formulated in Euclidean spacetime. 

This chapter introduces the theoretical foundations and current status of lattice-QCD calculations of PDFs, with emphasis on modern approaches based on spatially nonlocal matrix elements. We first review the light-cone definitions of quark and gluon PDFs, their Mellin moments, and the connection to QCD factorization. We then explain how large-momentum effective theory, short-distance factorization, and the short-distance operator product expansion make it possible to relate Euclidean lattice observables to light-cone partonic structure. Particular attention is given to the elements that have improved over the last five years: renormalization of Wilson-line operators, perturbative matching, finite-momentum and finite-distance effects, reconstruction of the $x$ dependence, and systematic uncertainties. 

We summarize selected lattice results for proton quark PDFs, pion and kaon PDFs, gluon PDFs, and twist-3 distributions, highlighting both recent progress and remaining challenges. As will be demonstrated, lattice QCD is moving from proof-of-principle calculations toward systematically improvable determinations that can complement experimental data and global QCD analyses in mapping the partonic structure of hadrons.

\end{abstract}

\section{Introduction} 
\label{sec:intro}

The proton and the neutron, collectively referred to as nucleons, are the building blocks of atomic nuclei and account for nearly all of the mass of visible matter in the Universe. Although they have been known for about a century, their internal structure remains far from fully understood. In the simplest quark-model picture, the proton consists of two up quarks and one down quark, $uud$, while the neutron consists of one up quark and two down quarks, $udd$; these are the nucleons' valence quarks. Such a simplistic picture is useful for classifying hadrons and understanding many of their quantum numbers, but it provides only a limited description of their internal dynamics.

Quarks interact through the strong force by exchanging gluons. A quark can radiate a gluon, a gluon can split into a quark-antiquark pair, and gluons can themselves radiate, interact, and recombine. The resulting sea of quarks and antiquarks includes not only the light up and down quarks, but also heavier flavors, such as strange and charm quarks. The interior of a nucleon is therefore not a static collection of three constituent particles, but a highly dynamical quantum system of interacting quarks, antiquarks, and gluons, collectively known as partons. Their motion and interactions ultimately give rise to macroscopic hadronic properties, such as the nucleon mass, spin, and electromagnetic structure. Separate chapters of this Encyclopedia can be found for the spin, that is ``Spin Structure of the Nucleon: Overview''~\cite{Hatta:2026txn} and ``Spin structure of the nucleon: experiment''. More details on the history of QCD  can be found in the chapter ``From the quark-parton model to QCD'' of this Encyclopedia~\cite{Soper:2026kvo} and the review article~\cite{Gross:2022hyw}. 

The strong interaction is described by quantum chromodynamics (QCD), the non-Abelian gauge theory of quarks and gluons, and is incorporated within the Standard Model. QCD is characterized by two seemingly contrasting properties. At high energies, or equivalently at short distances, the interaction becomes weak, allowing the use of perturbation theory. At the energy scale of hadrons, however, the interaction becomes strong, and quarks and gluons are confined inside color-neutral states. This coexistence of perturbative and nonperturbative regimes makes QCD exceptionally rich, but also makes quantitative studies of hadron structure particularly challenging. Nevertheless, significant progress has been achieved in recent years, with further advances expected from increasingly precise experiments, theoretical developments, and first-principles numerical calculations.

A central goal of modern nuclear and particle physics is, among others, to determine how the momentum and spin of a hadron are distributed among its quark and gluon constituents. This information is encoded in partonic correlation functions. Among these, the most fundamental ones are parton distribution functions (PDFs), which describe the longitudinal momentum and polarization structure of a hadron viewed at high energy. In an intuitive parton-model interpretation, a PDF quantifies the probability density for finding a quark, antiquark, or gluon carrying a fraction $x$ of the hadron's longitudinal momentum. More precisely, PDFs are matrix elements of gauge-invariant light-cone operators and depend not only on the momentum fraction $x$, but also on a scale, conventionally denoted by $\mu$. Their scale dependence is governed by perturbative QCD evolution, whereas their dependence on $x$ is intrinsically nonperturbative. More details can be found in the chapter ``Parton distributions of the proton and precision QCD'' in this Encyclopedia.

Different PDFs characterize complementary aspects of hadron structure. The unpolarized PDF describes the longitudinal momentum distribution of partons when their spin orientation is not resolved. The helicity PDF measures the difference between partons whose spin is aligned and anti-aligned with the longitudinal polarization of the parent hadron. The transversity PDF provides analogous information for transverse polarization. All the information from gluon- and flavor-dependent quark and antiquark distributions provides a one-dimensional picture of hadron structure in longitudinal momentum space. Together, these functions provide a one-dimensional picture of how the momentum and spin of a hadron are distributed among its partonic constituents.

PDFs are also the simplest members of a family of partonic distributions. Generalized parton distributions (GPDs) extend them to non-forward kinematics and correlate longitudinal momentum with spatial information. Transverse-momentum-dependent distributions retain information about intrinsic transverse momentum. In the forward limit, selected GPDs reduce to the familiar unpolarized, helicity, and transversity PDFs. The PDF and GPD chapters in this Encyclopedia are therefore complementary: the present chapter introduces the forward partonic structure of hadrons, while the companion chapter ``Generalized parton distribution functions from lattice QCD'' develops the more general non-forward framework. Information on GPDs from theory and experiment can be found in this Encyclopedia under chapters ``Generalized parton distributions'', and ``GPD phenomenology''. In addition, details on obtaining transerse-momentum dependent distributions from lattice QCD can be found in the chapter ``TMDs on the lattice'' in this Encyclopedia.

The first direct evidence for point-like constituents inside the nucleon emerged from the seminal deep-inelastic scattering (DIS) experiments performed at the Stanford Linear Accelerator Center in the late 1960s. These experiments played for the nucleon a role similar to that of Rutherford scattering for the atom: they revealed structure at distances much smaller than the overall size of the target. The approximate scaling of the measured structure functions with the Bjorken variable suggested that the scattering arose from nearly free, point-like constituents and led to the development of the parton model. The subsequent formulation of QCD identified these constituents with quarks and gluons and explained deviations from exact scaling through perturbative radiation and QCD evolution.

Deep-inelastic scattering is an inclusive process: the scattered lepton is detected, but the full hadronic final state is not reconstructed and, in particular, the fate of the initial hadron is not observed. From the theoretical point of view, the short-distance interaction can be treated perturbatively when the momentum transfer is sufficiently large, while the long-distance structure of the incoming hadron is encoded in PDFs. This separation is formalized through QCD factorization. We refer the reader to the chapters of this Encyclopedia ``QCD collinear factorization and the renormalization group'' and ``Deeply inelastic scattering experiments – overview '' for more details. Schematically, a measurable cross section is written as a convolution of perturbatively calculable coefficient functions with universal, nonperturbative PDFs, supplemented by corrections suppressed by powers of the hard scale, 
\begin{equation}
d\sigma
=
\sum_i
C_i(Q,\mu,\alpha_s(\mu))
\otimes
f_{i/H}(\mu)
+
\mathcal{O}\!\left(\frac{\Lambda_{\rm QCD}^p}{Q^p}\right).
\label{eq:factorization}
\end{equation}
The same PDFs can therefore be used across many processes, including DIS, Drell-Yan production, electroweak-boson production, heavy-flavor production, and jet observables.

Over several decades, increasingly precise experimental data from fixed-target facilities and high-energy colliders have been combined in global QCD analyses to determine PDFs. Such analyses not only simultaneously fit data from many processes, but they also evolve distributions between different scales using the Dokshitzer-Gribov-Lipatov-Altarelli-Parisi (DGLAP) equations, and estimate uncertainties arising from experimental input, parametrization choices, and theoretical assumptions. They have produced an increasingly detailed picture of the quark and gluon structure of the proton and are indispensable for predictions at hadron colliders; for a review, see, e.g., Ref.~\cite{Gao:2017yyd}. More details on the topic can be found in the chapter of this Encyclopedia ``Global analysis of PDs'' and ``Global analysis of helicity-dependent parton distribution functions''~\cite{Nocera:2026est}.  Nevertheless, substantial uncertainties remain. These are especially important for certain flavor combinations, polarized and transversity distributions, the gluon distribution in some kinematic regions, and the behavior of PDFs at very small or very large $x$. The structure of the pion and the kaon is considerably less constrained experimentally, because they cannot be used as stable targets and their PDF is considerably less constrained experimentally.

The importance of PDFs extends beyond their role as descriptors of hadron structure. They are essential inputs for precision predictions in high-energy experiments, including measurements of Standard Model processes and searches for physics beyond the Standard Model. Uncertainties in PDFs propagate directly to theoretical predictions for hadron-collider observables, making their accurate determination a central component of modern particle and nuclear phenomenology. 

Lattice QCD offers a complementary route to partonic structure directly from the underlying theory. In lattice QCD, the QCD path integral is formulated on a discrete Euclidean spacetime grid and evaluated numerically using Monte Carlo methods. In principle, this provides a systematically improvable, first-principles framework in which the effects of quarks and gluons are treated nonperturbatively. In practice, however, PDFs cannot be calculated directly from their defining light-cone correlations because standard lattice simulations are performed in Euclidean rather than Minkowski spacetime. Light-cone separations necessarily involve real time, whereas the Euclidean formulation replaces it with imaginary time to make the path integral numerically feasible. For many years, this incompatibility restricted lattice studies of PDFs primarily to their Mellin moments. Through the operator product expansion (OPE), moments of PDFs can be related to matrix elements of local twist-2 operators that are accessible in Euclidean lattice calculations. This ``traditional-method'' program has yielded important information on quark momentum fractions, axial and tensor charges, and low moments of quark and gluon PDFs and GPDs. Its reach is limited, however, because only the lowest few moments can be determined reliably. Higher moments suffer from rapidly increasing statistical noise and complicated operator mixing, making a model-independent reconstruction of the full $x$ dependence impractical.

A major development over the last decade has been the emergence of methods based on spatially nonlocal operators. These approaches compute equal-time Euclidean correlations in hadrons with finite momentum and relate them to light-cone PDFs via perturbative factorization. The two most widely used frameworks are large-momentum effective theory (LaMET), commonly associated with quasi-PDFs~\cite{Ji:2013dva,Ji:2014gla}, and short-distance factorization (SDF), often formulated in terms of pseudo-PDFs or Ioffe-time distributions~\cite{Radyushkin:2017cyf,Radyushkin:2020}. They use closely related lattice matrix elements but organize the matching and reconstruction differently. In the large-momentum approach, one typically reconstructs a momentum-space quasi-distribution and subsequently matches it to the light-cone PDF. In short-distance factorization, matching is performed first in coordinate space, followed by reconstruction of the $x$ dependence. Both methods are affected by finite-momentum, finite-distance, perturbative, and reconstruction uncertainties, and their complementarity provides valuable opportunities for consistency checks and combined analyses. The reader can find a separate chapter on LaMET in the chapter ``Large Momentum Effective Field Theory'' of this Encyclopedia.

It should be noted that the extraction of PDFs from such novel methods is a multistage problem. It requires the computation of boosted-hadron correlation functions, control of excited-state contamination, renormalization of nonlocal operators, perturbative matching to light-cone quantities, and reconstruction of a continuous function from data available only over a finite range of spatial separations and momenta. Each stage may introduce systematic uncertainties that must be quantified before lattice PDFs can achieve a fully controlled level of precision. At the same time, lattice QCD has distinct advantages: it can access flavor combinations that are difficult to isolate experimentally, provide information on moments and coordinate-space correlations, study polarized and gluonic structure, and make predictions on quantities for which experimental information is sparse.

Although LaMET and SDF are the most widely used modern approaches for accessing the $x$ dependence of PDFs from lattice QCD, they are part of a broader set of strategies developed to overcome the Euclidean light-cone obstruction. Other proposals include hadronic tensor methods~\cite{Liu:1993cv}, auxiliary-field or heavy-quark methods~\cite{Aglietti:1998ur,Detmold:2005gg,Braun:2007wv}, Compton amplitude~\cite{Chambers:2017dov}, current-current correlators~\cite{Ma:2014jla,Ma:2017pxb}, related formulations based on suitable Euclidean observables that can be factorized into light-cone PDFs~\cite{Braun:1994jq}, alternative approaches to access high moments~\cite{Davoudi:2012ya,Shindler:2023xpd}, and matrix elements of nonlocal operators at a fixed gauge~\cite{Gao:2023lny}. These approaches differ in their practical implementation, factorization structure, and systematic uncertainties, but they share the same goal: to connect first-principles Euclidean correlation functions to the light-front partonic structure of hadrons. More detailed accounts of the development of these methods and of lattice calculations of PDFs can be found in recent reviews~\cite{Cichy:2018mum,Ji:2020ect,Constantinou:2020pek,Constantinou:2020hdm,Cichy:2021lih,Cichy:2021ewm}.

This chapter reviews the theoretical foundations and current status of PDFs from lattice QCD. We begin by introducing the light-cone definitions of PDFs, followed by a discussion on extracting their $x$ dependence from nonlocal matrix elements. Particular attention is given to LaMET and SDF, including the operator product expansion of spatially separated nonlocal operators leading to Mellin moments, their perturbative matching to light-cone distributions or light-cone Ioffe-time distributions, renormalization, inverse reconstruction, and the control of systematic uncertainties. We then review selected results for nucleon, pion, kaon, polarized, transversity, and gluon distributions, and conclude with the prospects for integrating lattice information into global analyses and achieving a quantitatively controlled description of hadron structure.

\section{Definition of PDFs}
\label{sec:pdf-definitions}

Parton distribution functions are defined as hadronic matrix elements of gauge-invariant light-cone operators. They encode the nonperturbative structure of a hadron and enter experimentally measurable cross sections through QCD factorization (see, e.g., Eq.~\eqref{eq:factorization}). In the parton-model picture and at leading order, PDFs may be interpreted as probability densities for finding quarks, antiquarks, or gluons carrying a fraction $x$ of the longitudinal momentum of a fast-moving hadron. In addition to carrying $x$ as a variable, PDFs depend on a scale $\mu$ associated with the renormalization of the light-cone operators. In phenomenological applications, this scale is usually identified with the factorization scale that separates long-distance partonic structure from short-distance coefficient functions. PDFs are therefore defined only after specifying both a scale and a factorization/renormalization scheme, conventionally the $\overline{\rm MS}$ scheme.

For a hadron with four-momentum $P$ and spin $S$, the quark PDFs are obtained from bilocal quark operators separated along the light cone. Introducing light-cone coordinates $v^\pm=(v^0\pm v^3)/\sqrt{2}$, the unpolarized quark PDF (Dirac structure $\gamma^+$) is defined through
\begin{equation}
q(x,\mu)=
\int \frac{dz^-}{4\pi}\,
e^{ixP^+z^-}
\langle P,S|
\bar{\psi}_q(-z^-/2)\gamma^+
W(-z^-/2,z^-/2)
\psi_q(z^-/2)
|P,S\rangle_\mu ,
\label{eq:unpol-pdf-def}
\end{equation}
where $\psi_q$ is the quark field of flavor $q$, and $W(-z^-/2,z^-/2)$ is a Wilson line connecting the two quark fields along the light-cone direction~\footnote{The corresponding Euclidean matrix elements are usually written with one quark field at the origin and the other at a spatial separation $z$, connected by a straight Wilson line $W(0,z)$. This form is equivalent, by translation invariance, to the symmetric placement of the fields at $-z/2$ and $z/2$ used in the continuum light-cone definition, but it is more convenient for the construction of lattice correlators, as $z$ is an integer.}. The Wilson line is required for gauge invariance and is defined as the path-ordered exponential of the gluon field along the separation between the fields. The subscript $\mu$ indicates that the operator has been renormalized at scale $\mu$. The variable $x$ denotes the fraction of the hadron's longitudinal momentum carried by the struck parton. 
Similarly to the unpolarized case of $q$, the helicity ($\Delta q$) and transversity ($\delta q$) PDFs are defined by replacing the Dirac structure $\gamma^+$ in Eq.~\eqref{eq:unpol-pdf-def} with $\gamma^+\gamma_5$ and $i\sigma^{j+}\gamma_5$ ($j=1,2$), respectively. The helicity PDF is defined for a longitudinally polarized hadron, with spin vector $S_L$, and measures the difference between quarks with spin aligned and anti-aligned with the hadron momentum. The transversity PDF is defined for a transversely polarized hadron, with spin vector $S_T$, and describes the corresponding spin asymmetry for quarks polarized transverse to the hadron momentum. $\delta q$ is chiral-odd and therefore cannot be accessed in inclusive DIS alone; it requires another chiral-odd object in the process. Together, the unpolarized, helicity, and transversity PDFs provide the leading-twist~\footnote{The original definition of twist, often called geometric twist, is $\tau=d-s$, where $d$ is the canonical mass dimension of the operator and $s$ is its Lorentz spin; for local symmetric and traceless operators, $s$ is associated with the number of Lorentz indices~\cite{Gross:1971wn,Brandt:1970kg}. For PDFs, one often uses collinear twist, based on light-cone power counting and directly related to suppression by powers of the hard scale~\cite{Kogut:1969xa,Jaffe:1991ra}. More details can be found in Ref.~\cite{Braun:2022gzl}.} (twist-2) quark PDFs for a spin-$1/2$ hadron. 

The definitions above describe quark distributions, in which the nonlocal light-cone operator is built from quark fields. An analogous construction applies to gluons, but the operator is built from gluon field-strength tensors rather than quark bilinears. This difference makes the gluon case technically more involved, but the physical interpretation is parallel. A common convention for the unpolarized gluon distribution is
\begin{equation}
xg(x,\mu)=
\frac{1}{P^+}
\int \frac{dz^-}{2\pi}\,
e^{ixP^+z^-}
\langle P|
F^{+\alpha}(-z^-/2)
W(-z^-/2,z^-/2)
F^{+}_{\ \alpha}(z^-/2)
|P\rangle_\mu ,
\label{eq:gluon-pdf-def}
\end{equation}
with an appropriate Wilson line in the adjoint representation. Here, $F^{\mu\nu}=F_a^{\mu\nu}t^a$ is the gluon field-strength tensor, with
\begin{equation}
F_a^{\mu\nu}
=
\partial^\mu A_a^\nu-\partial^\nu A_a^\mu
+g f^{abc} A_b^\mu A_c^\nu ,
\end{equation}
where $A_a^\mu$ is the gluon gauge field, $f^{abc}$ are the SU(3) structure constants, and $a=1,\ldots,8$ is an adjoint color index. The unpolarized gluon PDF, $g(x,\mu)$, has a partonic interpretation analogous to that of the unpolarized quark PDF: it describes the distribution of gluons carrying a longitudinal momentum fraction $x$ of the parent hadron, irrespective of their polarization. The gluon helicity PDF, $\Delta g$, is obtained from the corresponding polarized gluonic operator, and measures the difference between gluons with helicity aligned and anti-aligned with the longitudinal polarization of the parent hadron. We note that there is no leading-twist gluon transversity distribution for a spin-$1/2$ hadron such as the proton.

The support of PDFs is restricted to the interval $-1\leq x\leq1$, and, in the usual convention, positive $x$ corresponds to quarks and negative $x$ encodes antiquark information. Equivalently, one often works with separate quark and antiquark PDFs defined in the region $0\leq x\leq1$. In such a case, the crossing relations are
\begin{equation}
q(-x,\mu)=-\bar{q}(x,\mu),\qquad
\Delta q(-x,\mu)=\Delta \bar{q}(x,\mu),\qquad
\delta q(-x,\mu)=-\delta \bar{q}(x,\mu).
\label{eq:negative-x-relations}
\end{equation}

It is often useful to form valence, sea, non-singlet, and singlet combinations. For example, the valence distribution is
\begin{equation}
q_v(x,\mu)=q(x,\mu)-\bar{q}(x,\mu),
\label{eq:valence-def}
\end{equation}
while the quark singlet distribution is
\begin{equation}
\Sigma(x,\mu)=\sum_q \left[q(x,\mu)+\bar{q}(x,\mu)\right].
\label{eq:singlet-def}
\end{equation}
Non-singlet combinations, such as $u-d$, are often simpler theoretically and computationally because they do not mix with gluon operators and, in lattice QCD, may avoid quark-disconnected contributions. Singlet and gluon distributions, by contrast, are coupled under QCD evolution and require the inclusion of quark-disconnected diagrams and treatment of the quark-gluon mixing, which is also present in the continuum.

As mentioned in the introduction, the light-cone nature of Eqs.~\eqref{eq:unpol-pdf-def} - \eqref{eq:gluon-pdf-def} is the main difficulty for lattice QCD. Standard lattice calculations are performed in Euclidean spacetime and therefore cannot evaluate light-cone correlations directly. A way to extract information on PDFs is through the traditional computation of Mellin moments using local operators. These moments provide important integral constraints and connect the light-cone formalism to local operators. Using as an example the unpolarized quark PDF, its Mellin moments are defined by
\begin{equation}
\langle x^n\rangle_q(\mu)=
\int_{-1}^{1} dx\, x^n q(x,\mu),
\label{eq:mellin-moment}
\end{equation}
or, equivalently, by integrals over $0\leq x\leq1$ involving quark and antiquark combinations. Through the operator product expansion (OPE), these moments are related to matrix elements of local twist-2 operators,
\begin{equation}
\mathcal{O}^{\{\mu_0\cdots\mu_n\}}_q
=
\bar{\psi}_q\gamma^{\{\mu_0}
iD^{\mu_1}\cdots iD^{\mu_n\}}
\psi_q
-\mathrm{traces},
\label{eq:twist-two-op}
\end{equation}
where curly brackets denote symmetrization over Lorentz indices and subtraction of traces. Analogous local operators exist for helicity, transversity, and gluon moments. This relation between moments and local operators was historically the primary way in which lattice QCD accessed PDF information.

Several moments have direct physical interpretations. The zeroth moment of the valence unpolarized quark distribution gives the number of valence quarks,
\begin{equation}
\int_0^1 dx\, q_v(x,\mu)=N_q ,
\label{eq:valence-sum-rule}
\end{equation}
where $N_q$ is the number of valence quarks of flavor $q$ in the hadron. The first moments of quark and gluon momentum distributions satisfy the momentum sum rule,
\begin{equation}
\sum_q \int_0^1 dx\, x\left[q(x,\mu)+\bar{q}(x,\mu)\right]
+
\int_0^1 dx\, xg(x,\mu)
=1.
\label{eq:momentum-sum-rule}
\end{equation}
Similarly, the first moment of the helicity distribution contributes to the quark spin decomposition of the nucleon, while the first moment of the transversity distribution gives the tensor charge,
\begin{equation}
g_T^q(\mu)=
\int_0^1 dx\,
\left[h_1^q(x,\mu)-h_1^{\bar q}(x,\mu)\right].
\label{eq:tensor-charge}
\end{equation}
These sum rules and moments provide important benchmarks for both phenomenological extractions and lattice-QCD calculations.

The following section focuses on modern lattice approaches based on spatially nonlocal matrix elements, which provide access to the $x$ dependence of PDFs through perturbative factorization, as well as on the use of the short-distance OPE applied directly to such nonlocal matrix elements to extract Mellin moments. Traditional calculations of Mellin moments from local twist-2 operators play an important role in lattice studies of hadron structure, but a detailed discussion of that program is beyond the scope of this chapter, and we refer the reader to the chapter ``Lattice QCD for nucleon form factors'', as well as ``Nucleon and nuclear form factors: theory and experiment''.

\section{Accessing PDFs from Euclidean lattice QCD}
\label{sec:lattice-access}

The light-cone definitions introduced in Sec.~\ref{sec:pdf-definitions} provide the natural theoretical formulation of PDFs in continuum QCD. However, they also reveal the central difficulty for lattice QCD. PDFs are defined through correlations of fields separated along the light cone, while lattice-QCD calculations are performed in Euclidean spacetime. This mismatch prevents a direct evaluation of the light-cone operators that define PDFs. The modern approaches mentioned in Sec.~\ref{sec:intro} are designed for lattice QCD to overcome this challenge by computing Euclidean matrix elements of spatially separated operators and relating them to light-cone PDFs through perturbative factorization (matching formalism).

Focusing on the methodologies of LaMET and SDF, the starting point is a Euclidean hadronic matrix element of a spatially nonlocal quark bilinear,
\begin{equation}
M_\Gamma^q(z,P_3)
=
\langle P|
\bar{\psi}_q(0)\Gamma W(0,z)\psi_q(z)
|P\rangle,
\label{eq:nonlocal-me}
\end{equation}
where the Wilson line extends along a spatial direction, conventionally chosen to be the third direction, and the hadron carries momentum $P_3$ along the same direction. The Dirac matrix $\Gamma$ selects the desired PDF channel. For example, $\Gamma=\gamma^0$ or $\gamma^3$ may be used for the unpolarized distribution, $\Gamma=\gamma^3\gamma_5$ for helicity, and $\Gamma=i\sigma^{j3}\gamma_5$ for transversity. In practical lattice calculations, the choice of $\Gamma$ is also influenced by renormalization and mixing properties at finite lattice spacing. For Wilson-type fermions, for instance, the use of $\gamma^3$ in the unpolarized case can induce mixing with the scalar operator (in physical basis), while $\gamma^0$ avoids this mixing~\cite{Constantinou:2017sej}.

The matrix element in Eq.~\eqref{eq:nonlocal-me} depends on the spatial separation $z$, the hadron momentum $P_3$. It is often useful to express this dependence in terms of the Ioffe time,
\begin{equation}
\nu = zP_3 ,
\label{eq:ioffe-time}
\end{equation}
which is the Euclidean analog of the light-cone distance conjugate to the partonic momentum fraction $x$. The same underlying matrix elements can be analyzed in different but related ways: large-momentum effective theory (LaMET), which leads to quasi-PDFs; and short-distance factorization (SDF), which is commonly formulated in terms of pseudo-PDFs or Ioffe-time distributions. In addition, the short-distance operator product expansion of the same nonlocal matrix elements can be used to extract Mellin moments via an OPE without reconstructing the full $x$ dependence.

The nonlocal matrix elements must be renormalized before they can be related to light-cone PDFs, which introduces the scale $\mu$. The Wilson line in Eq.~\eqref{eq:nonlocal-me} introduces additional ultraviolet divergences beyond those of local operators. In particular, straight Wilson-line operators contain a linear divergence proportional to the length of the Wilson line~\cite{Mandelstam:1968hz,Polyakov:1979gp,Dotsenko:1979wb,Brandt:1981kf}, in addition to logarithmic divergences. Several renormalization prescriptions have been proposed for such operators, including a few variations of regularization-independent momentum-subtraction schemes~\cite{Constantinou:2017sej,Alexandrou:2017huk,Chen:2017mzz,Zhang:2020rsx,Constantinou:2022aij}, ratio schemes~\cite{Radyushkin:2017cyf,Orginos:2017kos}, hybrid schemes~\cite{Ji:2020brr}, and self-renormalization procedures~\cite{LatticePartonLPC:2021gpi}. These prescriptions remove the same Wilson-line ultraviolet divergences, but differ in whether the renormalization factors are obtained from off-shell quark matrix elements, ratios of hadronic matrix elements, short- and long-distance separation, or fits to the nonlocal matrix elements themselves. 

Once this ultraviolet renormalization has been carried out, the remaining step is to relate the renormalized Euclidean matrix elements to light-cone PDFs in a continuum scheme, most often $\overline{\rm MS}$. This is the scheme commonly used in global analyses of experimental data, and it therefore provides the natural basis for comparing lattice-extracted PDFs with phenomenological determinations. The relation to the light-cone PDF is implemented differently in LaMET and SDF. In the LaMET approach, the spatial matrix element is first Fourier transformed with respect to the separation $z$ to define a quasi-PDF,
\begin{equation}
\tilde{q}(x,P_3,\mu)
=
\int_{-\infty}^{\infty}
\frac{dz}{4\pi}\,
e^{ixP_3 z}
M_\Gamma^q(z,P_3,\mu).
\label{eq:quasi-pdf}
\end{equation}
The quasi-PDF is not identical to the light-cone PDF at finite momentum. However, for sufficiently large hadron momentum, the quasi-PDF and the light-cone PDF have the same infrared physics and differ only in ultraviolet contributions that can be calculated perturbatively~\cite{Ji:2015jwa,Chen:2016fxx,Ji:2017oey,Ishikawa:2017faj,Li:2018tpe}. The same arguments hold for the pseudo-PDF in the SDF approach. This leads to a factorization relation of the form
\begin{equation}
\tilde{q}(x,P_3,\mu)
=
\int_{-1}^{1}
\frac{dy}{|y|}
C_{\rm LaMET}\left(\frac{x}{y},\frac{\mu}{P_3}\right)
q(y,\mu)
+
\mathcal{O}\left(\frac{\Lambda_{\rm QCD}^2}{P_3^2},
\frac{M^2}{P_3^2}\right),
\label{eq:lamet-factorization}
\end{equation}
where $C_{\rm LaMET}$ is a perturbatively calculable matching kernel, $M$ is the hadron mass, and the power corrections arise from finite-momentum and higher-twist effects. In the infinite-momentum limit, the quasi-PDF approaches the light-cone PDF. At finite momentum, the reliability of the extraction depends on controlling the momentum dependence, perturbative matching, higher-twist effects, and the finite range of $z$ available in the lattice calculation, which impacts the effectiveness of the reconstruction.

In the SDF approach, one instead works in coordinate space. The matrix element is written as a function of the Ioffe time $\nu=zP_3$ and the invariant distance $z^2$. A commonly used object is the reduced Ioffe-time distribution, in which ratios of matrix elements are formed to cancel the Wilson-line renormalization and reduce certain systematic effects. For sufficiently small spatial separations, the Euclidean Ioffe-time distribution can be factorized into the light-cone Ioffe-time distribution,
\begin{equation}
\mathcal{M}(\nu,z^2,\mu)
=
\int_{-1}^{1} dx\,
e^{ix\nu}
q(x,\mu)
+
\mathrm{perturbative\ corrections}
+
\mathcal{O}(z^2\Lambda_{\rm QCD}^2).
\label{eq:itd-lightcone}
\end{equation}
More explicitly, the short-distance factorization relation may be written as
\begin{equation}
\mathcal{M}(\nu,z^2)
=
\int_{-1}^{1} dy\,
C_{\rm SDF}(y,z^2\mu^2)
\mathcal{Q}(y\nu,\mu)
+
\mathcal{O}(z^2\Lambda_{\rm QCD}^2),
\label{eq:sdf-factorization}
\end{equation}
where $\mathcal{Q}(\nu,\mu)$ is the light-cone Ioffe-time distribution, related to the PDF by Fourier transformation, 
\begin{equation}
\mathcal{Q}(\nu,\mu)
=
\int_{-1}^{1} dx\, e^{ix\nu} q(x,\mu).
\label{eq:lc-itd-pdf}
\end{equation}
In this approach, the matching is performed in coordinate space before reconstructing the $x$ dependence. The leading power corrections are controlled by the smallness of $z^2$, rather than directly by the inverse hadron momentum.

It should be noted that the real and imaginary parts of the coordinate-space matrix elements provide complementary information. For the unpolarized quark PDF, the real part is associated with the valence distribution (Eq.~\eqref{eq:valence-def}), while the imaginary part contains information on combinations involving sea quarks (Eq.~\eqref{eq:singlet-def}). Similar relations hold for helicity and transversity distributions. Since the imaginary part is often noisier, the separation of quark and antiquark distributions is generally more challenging than the determination of valence-like combinations. This is one reason why lattice determinations of nonsinglet or valence PDFs are typically more mature than determinations of sea-quark distributions. Another reason is that singlet distributions require the calculation of disconnected diagrams, which are very noisy and challenging to compute. We refer to Sec.~\ref{subsec:results-quark-proton} for further discussion.

The LaMET and SDF approaches are closely related because they can be applied to the same Euclidean matrix elements. Their main practical difference lies in the order in which matching and Fourier reconstruction are performed. In LaMET, one typically reconstructs a quasi-PDF in momentum space and then matches it to the light-cone PDF. In SDF, one first matches the coordinate-space matrix element or Ioffe-time distribution and then reconstructs the PDF. In the formal limits of large momentum and short distance, the two approaches are expected to yield the same light-cone distribution. However, at the lattice spacings, momenta, and Wilson-line lengths available in practice, they may have different sensitivities to systematic effects. Comparing the two approaches on the same matrix elements, therefore, provides a valuable test of methodology-related systematic uncertainties. Another difference in the analysis is that the matrix elements of each momentum boost are analyzed individually, whereas in SDF, several momenta enter the same analysis.

In addition to extracting the $x$ dependence of PDFs, the same spatially nonlocal matrix elements can also be used to extract Mellin moments through a short-distance OPE. One expands the nonlocal operator at small $z$ in terms of local twist-2 operators,
\begin{equation}
\bar{\psi}(0)\Gamma W(0,z)\psi(z)
=
\sum_{n=0}^{\infty}
C_n(z^2\mu^2)\,
z_{\mu_1}\cdots z_{\mu_n}
\mathcal{O}^{\mu_1\cdots\mu_n}(\mu)
+
\mathrm{higher\ twist},
\label{eq:nonlocal-ope}
\end{equation}
where the Wilson coefficients $C_n$ are perturbatively calculable, and the matrix elements of the resulting local operators are proportional to Mellin moments of PDFs. This strategy has the advantage of avoiding the ill-posed inverse problem associated with reconstructing a continuous function of $x$ from a finite range of Ioffe times. It also provides an alternative way to determine moments beyond the traditional local-operator approach, although it still requires control over OPE convergence, perturbative truncation, scale evolution, and higher-twist effects.

To summarize, these methods share several common sources of systematic uncertainties. The lattice matrix elements are affected by statistical noise, finite-volume effects, discretization effects, unphysical quark masses in some calculations, and excited-state contamination. The use of boosted hadrons introduces additional challenges because the signal-to-noise ratio deteriorates as the hadron momentum increases. The nonlocal operator requires careful renormalization, and the perturbative matching introduces dependence on the order of perturbation theory and on the chosen renormalization scheme. Finally, the finite range of accessible spatial separations and hadron momenta limits the available Ioffe-time coverage, making the reconstruction of the $x$ dependence a nontrivial inverse problem.

Despite these challenges, Euclidean matrix elements of nonlocal operators have transformed the role of lattice QCD in PDF studies. They provide access not only to a few low moments, but also to the functional dependence of PDFs on $x$, to Ioffe-time distributions, and to complementary moment extractions through the nonlocal OPE. The next section briefly discusses selected theoretical improvements to these approaches and the reconstruction strategies used to obtain PDFs from finite-lattice data.

\section{Improving determinations of PDFs}
\label{sec:systematics}

The theoretical development of lattice methods for PDFs has focused on turning spatially nonlocal Euclidean matrix elements into quantitatively controlled, factorizable observables. A first key advance was the understanding of the renormalization properties of straight-Wilson-line operators. Such operators contain a Wilson-line power divergence, analogous to a static-quark mass divergence, as well as logarithmic ultraviolet divergences. Theoretical studies showed that, after subtraction of the Wilson-line mass counterterm, the relevant nonlocal quark operators are, in the continuum, multiplicatively renormalizable to all orders in perturbation theory~\cite{Ji:2015jwa,Chen:2016fxx,Ji:2017oey,Ishikawa:2017faj}. The corresponding renormalizability arguments have also been extended to quasi-gluon operators~\cite{Li:2018tpe}.

A second major development was the derivation and improvement of perturbative matching kernels. In LaMET, the matching relates quasi-PDFs at finite hadron momentum to light-cone PDFs, while in SDF it relates Euclidean Ioffe-time distributions at short spatial separations to light-cone Ioffe-time distributions or PDFs. In both cases, the matching kernels remove the ultraviolet difference between the Euclidean and light-cone quantities while preserving the common infrared physics. One-loop matching established the practical connection between quark quasi-PDFs and light-cone PDFs~\cite{Xiong:2013bka}, while the corresponding SDF/pseudo-PDF matching was developed in coordinate space using Ioffe-time distributions and reduced Ioffe-time distributions~\cite{Radyushkin:2017cyf,Orginos:2017kos,Radyushkin:2018cvn,Zhang:2018ggy}. Subsequent work clarified factorization, scheme conversion, and matching in lattice renormalization schemes~\cite{Izubuchi:2018srq}. The matching framework has also been extended to singlet quark and gluon channels, where quark-gluon mixing must be included~\cite{Wang:2019tgg,Yao:2022vtp}, and to gluon pseudo-distributions in the SDF approach~\cite{Balitsky:2019krf,Balitsky:2021bds}. More recently, matching coefficients have been pushed to NNLO accuracy for nonsinglet quark distributions~\cite{Li:2020xml,Chen:2020ody}, reducing perturbative truncation uncertainties in modern PDF extractions. Matching kernels appropriate to hybrid renormalization schemes have also been derived for nonsinglet quark PDFs and for singlet quark and gluon quasi-PDFs~\cite{Chou:2022drv,Chen:2024jkb}. Finally, related matching and factorization formulae have been developed for twist-3 quasi- and pseudo-distributions~\cite{Bhattacharya:2020xlt,Bhattacharya:2020jfj,Braun:2021aon,Braun:2021gvv}, extending the nonlocal-operator program beyond leading twist.

The perturbative matching is closely connected to the choice of renormalization prescription. The UV divergence must be removed nonperturbatively or through an equivalent renormalization prescription before the matrix elements can be matched to light-cone PDFs.  Several prescriptions have been developed in the last decade. Regularization-independent momentum-subtraction schemes were the first to be proposed for quasi-PDFs~\cite{Constantinou:2017sej,Alexandrou:2017huk,Chen:2017mzz,Zhang:2020rsx,Constantinou:2022aij}. Ratio schemes exploit cancellations of Wilson-line divergences in suitable ratios of matrix elements~\cite{Radyushkin:2017cyf,Orginos:2017kos}. More recently, hybrid~\cite{Ji:2020brr} and self-renormalization~\cite{LatticePartonLPC:2021gpi} schemes separate the short- and long-distance behavior of the matrix element and are designed to improve the stability of the renormalized result over the range of Wilson-line lengths used in practice. 

Finite hadron momentum is one of the defining variables of LaMET calculations, and has been a major challenge since the first calculations. The formal connection between quasi-PDFs and light-cone PDFs becomes exact only in the infinite-momentum limit. At the momenta accessible in lattice QCD, the extracted distributions receive corrections suppressed by powers of the hadron momentum, such as $\Lambda_{\rm QCD}^2/P_3^2$ and $M^2/P_3^2$. The latter are often referred to as target-mass corrections and can be treated analytically in certain formulations. More general finite-momentum effects are associated with higher-twist contributions and are more difficult to remove. Reaching sufficiently large boosts is therefore essential, but in practice, it is limited by the rapid degradation of the signal-to-noise ratio and by increased excited-state contamination. Momentum smearing~\cite{Bali:2016lva} has been a crucial addition in this respect, as it improves the overlap of interpolating fields with boosted hadron states and makes matrix elements at larger momenta statistically accessible~\cite{Alexandrou:2016jqi}. A practical strategy is to calculate matrix elements at several values of $P_3$ and study the stability of the final result under changes in the boost. Agreement among different boosts after matching provides evidence that residual power corrections are under control.

In SDF, the analogous expansion parameter is the spatial separation $z$. The factorization is valid at short distances, where $z^2\Lambda_{\rm QCD}^2$ is small, and the nonlocal operator can be expanded perturbatively. At larger values of $z$, higher-twist effects become more important. Thus, SDF analyses require careful selection of the $z$ range used in the fit or matching procedure. The useful window must be large enough to provide sensitivity to the Ioffe-time dependence, but small enough that the short-distance expansion remains reliable. This balance is one of the main practical challenges in coordinate-space approaches. The short-distance OPE of nonlocal matrix elements provides an additional development that connects the modern nonlocal-operator program with the traditional moment program. By expanding the nonlocal operator around small $z$, one obtains a tower of local twist-2 operators whose matrix elements are proportional to Mellin moments of PDFs. In this approach, the Wilson coefficients are perturbatively calculable, while the moments are extracted from the dependence of the lattice matrix elements on $z$ and $P_3$. 

Power corrections and renormalon effects have also received increasing attention for quasi- and pseudo-distributions \cite{Braun:2018brg,Zhang:2023bxs,Holligan:2024wpv}.
The separation between perturbative Wilson coefficients and nonperturbative matrix elements is not unique beyond a given perturbative order, and factorial growth in perturbative expansions can lead to ambiguities that must be compensated by power-suppressed terms. These issues are particularly relevant for Wilson-line operators, where linear divergences, long-distance behavior, and perturbative matching are interconnected. Improved treatments of renormalization and matching aim to reduce such ambiguities and to make the separation between perturbative and nonperturbative physics more stable.

Discretization effects are another important area of improvement. The nonlocal operator has a spatial direction, and large hadron momenta can enhance cutoff effects. In addition, such matrix elements contain ${\cal O}(a)$ effects, unlike local operators, which can be ${\cal O}(a)$ improved ($a$: lattice spacing). Therefore, having calculations at multiple lattice spacings is even more crucial for such matrix elements. It should also be noted that the value of $a$ of a given ensemble controls the physical values of $z$ and $P_3$, which are essential for SDF/OPE (LaMET). A related development is the use of kinematically enhanced interpolating operators for boosted hadrons~\cite{Zhang:2025hyo,Reitinger:2026hta}. These operators are designed to improve the overlap with the boosted ground state, and hence the signal-to-noise ratio, at large momenta. 
Excited-state contamination may also be enhanced, as boosted hadrons are more difficult to isolate than hadrons at rest because the energy spectrum becomes denser. Multi-state fits, calculations at several source-sink separations, optimized interpolating fields, and momentum smearing are commonly used to improve ground-state isolation. 

Overall, the theoretical development of lattice methods for PDFs has moved the field from exploratory calculations toward systematically improvable determinations. The key advances include higher-order matching, improved renormalization schemes, better understanding of power corrections, use of the nonlocal OPE for Mellin moments, control of operator mixing, and more robust treatments of excited states and discretization effects. These improvements are essential because the final uncertainty of a lattice PDF needs to incorporate systematic effects. The next section addresses the remaining step: the reconstruction of the $x$ dependence from the finite and noisy lattice information.

\section{Reconstruction of the $x$ dependence}
\label{sec:x-reconstruction}

The factorization relations discussed in Sec.~\ref{sec:lattice-access} formally connect Euclidean lattice matrix elements to light-cone PDFs. However, they do not by themselves solve the practical problem of determining the distribution functions, which are continuous functions of $x$. This problem is intrinsically ill-posed for both LaMET and SDF, even though it manifests in different steps of the analysis. It is useful to distinguish between reconstructions performed in momentum space and those performed in coordinate space. We remind the reader that LaMET requires a momentum-space strategy, and one first obtains a quasi-PDF through a Fourier transform and then applies perturbative matching to the light-cone PDF. In a coordinate-space analysis, that is SDF approach, one first applies the matching relation and then reconstructs the PDF. In principle, both routes should lead to the same light-cone distribution once all systematic effects are controlled. In practice, however, they may respond differently to systematic uncertainties. 

In terms of the inverse problem, it manifests as follows. The matrix elements entering Eq.~\eqref{eq:quasi-pdf} span over a finite set of spatial separations, and, in LaMET, the transform is effectively truncated. This truncation can lead to unphysical oscillations and distortions in the reconstructed $x$ dependence. Equivalently, in SDF, the Fourier transform of Eq.~\eqref{eq:lc-itd-pdf} requires knowledge of $\mathcal{Q}(\nu,\mu)$ over an infinite range of $\nu$. In lattice QCD, however, the maximum Ioffe time is limited by the largest spatial separation and hadron momentum for which the matrix element can be reliably computed. The purpose of this section is to summarize strategies developed recently to address this inverse problem, including large-$z$ extrapolations~\cite{Ji:2026vir}, Backus-Gilbert reconstruction~\cite{BackusGilbert}, parametric fits, nonparametric Bayesian approaches~\cite{Karpie:2019eiq} including a Gaussian process regression~\cite{Alexandrou:2020tqq,Medrano:2025cmg,medrano2026normalizingflowsreconstructpseudopdfs}, and artificial neural networks~\cite{Karpie:2019eiq,Cichy:2019ebf,Chu:2025jsi}. These methods provide practical ways to extract information about the $x$ dependence from finite lattice data, but they may also introduce additional assumptions, such as choices of functional form, priors, extrapolation behavior, or resolution criteria. These choices can affect the reconstructed PDFs and should therefore be included in the systematic uncertainty.

A large-$z$ extrapolation is a way to supplement the finite range of the lattice data with an assumed functional form for the long-distance behavior before performing the Fourier transform~\cite{Ji:2026vir}. This procedure can reduce artifacts associated with truncating data at some $z_{\rm max}$ and improve the stability of the reconstructed distribution. However, the extrapolated region is not directly constrained by lattice data and can influence the small-$x$ and large-$x$ behavior of the final PDF. The dependence on the chosen extrapolation form must therefore be treated as a source of systematic uncertainty.

Parametric fits provide a complementary approach. In this case, one assumes a functional form for the PDF, typically inspired by phenomenological parametrizations, and determines its parameters by fitting the corresponding lattice matrix elements or matched Ioffe-time distributions. An example is a form proportional to
\begin{equation}
q(x,\mu) = N x^\alpha (1-x)^\beta \left(1+c_1\sqrt{x}+c_2 x\right), \qquad 0<x<1,
\label{eq:parametric-pdf}
\end{equation}
possibly supplemented by constraints from normalization, Mellin moments, or positivity, where applicable. Fitting directly to coordinate-space matrix elements has the advantage that the finite range of lattice data is handled at the level of the observable that is actually computed. The drawback is parametrization bias: if the chosen ansatz is too restrictive (e.g., $c_1=c_2=0$), the resulting PDF may appear more precise than justified by the lattice data. This can be tested by varying the functional form, increasing the number of parameters, changing the fitted range in $z$ or $\nu$, and comparing with alternative reconstruction methods.

Another strategy is the Backus-Gilbert (BG) reconstruction method~\cite{BackusGilbert}, a model-independent technique designed to address the inverse problem without assuming a specific functional form for the PDF. The method constructs, independently for each value of $x$, an estimator of the PDF as a linear combination of the available coordinate-space or Ioffe-time data. The reconstructed quantity is therefore an estimate of the PDF localized around a chosen value of $x$, rather than the exact PDF at that point. The BG method chooses the linear combination of lattice data so that this localization is as sharp as possible while keeping the statistical uncertainty under control. In this way, the BG method provides a stable reconstruction while clearly showing the limitations imposed by the finite and noisy lattice data. Its main advantage is that it does not require a strong assumption about the functional form of the PDF. Its main limitation is that the reconstructed value at a given $x$ may still receive contributions from a relatively broad range of nearby $x$ values, especially where the lattice data are less constraining. 

Bayesian approaches provide a systematic way to include prior information while propagating uncertainties~\cite{Karpie:2019eiq,Alexandrou:2020tqq,Medrano:2025cmg,medrano2026normalizingflowsreconstructpseudopdfs}. Priors may be used to impose smoothness, large-$x$ behavior, positivity, or consistency with sum rules. Nonparametric Bayesian methods, such as Gaussian process regression, are useful because they do not require a fixed functional form for the PDF. Instead, they define a probability distribution over possible functions, with correlations between different values of $x$ controlled by a kernel. This can provide a flexible reconstruction while allowing the uncertainty to grow in regions where the lattice data provide little constraint. As with all Bayesian methods, however, the result can depend on the choice of prior or kernel, and it is important to test the stability of the reconstruction under reasonable variations of the prior assumptions.

Artificial neural networks offer a more flexible strategy for reconstructing PDFs~\cite{Karpie:2019eiq,Cichy:2019ebf,Chu:2025jsi}. In this approach, the PDF is represented by a neural network whose parameters are determined by training on the lattice data, typically through the corresponding coordinate-space matrix elements or matched Ioffe-time distributions. Neural networks can reduce the bias associated with fixed analytic parametrizations and can be especially useful when combining data from several momenta, multiple Wilson-line separations, or different factorization frameworks~\cite{Chu:2025jsi}. Physical constraints, such as support, normalization, small- and large-$x$ behavior can be incorporated into the architecture or the loss function. The flexibility of neural networks is both an advantage and a challenge: without sufficient validation, the network may learn features driven by noise, training choices, or insufficiently constrained regions of the data. 

Several consistency checks are important in any reconstruction of the $x$ dependence. The final PDF should be tested under variations of the maximum Wilson-line length, the range of Ioffe times included in the analysis, the hadron momenta used, and the treatment of the large-$z$ or large-$\nu$ behavior. It should also be tested against changes in the renormalization prescription, matching scale, perturbative order, and reconstruction method. Sum rules provide additional constraints: moments of the reconstructed PDF can be compared with known charges or independently determined momentum fractions, when available. Such comparisons help determine whether the result is driven primarily by the lattice data or by assumptions introduced in the reconstruction. In summary, the reconstruction of PDFs from lattice QCD is not a simple numerical Fourier transform, but an essential part of the calculation that has been improved over the years. Each method has advantages and limitations, and comparisons among them are important for assessing reconstruction-related uncertainties. As lattice data improve, robust reconstruction strategies will remain crucial for converting Euclidean matrix elements into reliable light-cone parton distributions.

\section{Current landscape of lattice PDFs}
\label{sec:lattice-results}

The methods described in the previous sections have been applied to a growing set of partonic observables, including unpolarized, helicity, transversity, and gluon PDFs, as well as the partonic structure of mesons such as the pion and kaon. The purpose of this section is not to provide an exhaustive list of all lattice calculations, but rather to summarize the main directions of progress, the quantities that are currently most mature, and the systematic limitations that remain. In reviewing the current status, we focus on the most advanced applications of LaMET and SDF, the two approaches most widely used in lattice calculations of $x$-dependent PDFs. Due to space limitations, the discussion is not intended to be exhaustive. Earlier exploratory calculations are mentioned but the discussion focuses on studies within the last five years, in which the analysis strategies, renormalization procedures, matching, reconstruction methods, and control of lattice systematics have been substantially improved. We also note that other strategies have led to lattice implementations or exploratory studies of PDF-related observables. These include current-current correlation and lattice-cross-section approaches applied to the pion and proton~\cite{Sufian:2019bol,Sufian:2020vzb,Zimmermann:2024zde,Zhang:2026lle}, gradient-flow-based methods applied to pion PDFs and their moments~\cite{Francis:2025rya,Francis:2025pgf}, and Compton-amplitude calculations of quantities related to inclusive structure functions~\cite{Can:2025jzf}. 

\subsection{Quark PDFs for the proton}
\label{subsec:results-quark-proton}

Quark PDFs of the proton were the first $x$-dependent PDFs to be studied extensively in lattice QCD. This was a natural starting point, because quark matrix elements have a better signal-to-noise ratio than gluonic matrix elements, and the nonlocal quark-bilinear operators have a simpler structure than their gluonic counterparts. The earliest lattice calculations focused primarily on the isovector flavor combination $u-d$, for which disconnected diagrams cancel in the isospin-symmetric limit. This considerably reduces the numerical cost and avoids the additional complication of mixing associated with the quark-singlet and gluon PDFs, which requires a dedicated matching formalism. 

The first stage of lattice calculations of nucleon quark PDFs was within the LaMET framework. These studies demonstrated that the calculation of boosted-hadron matrix elements of spatially nonlocal quark bilinears is feasible~\cite{Lin:2014zya,Alexandrou:2015rja}, that led to the systematic pursue of $x$-dependent calculations first at heavier pion masses~\cite{Lin:2014zya,Alexandrou:2015rja,Alexandrou:2016jqi,Chen:2016utp,Chen:2017mzz,Alexandrou:2018eet,Lin:2018pvv,Fan:2020nzz} and onward at physical pion mass for the unpolarized~\cite{Lin:2017ani,Alexandrou:2017dzj,
Chen:2018xof,Alexandrou:2018pbm,
Alexandrou:2018own,Alexandrou:2019lfo,
Bhat:2020ktg,Lin:2020fsj,
Gao:2022uhg}, helicity~\cite{Lin:2017ani,Alexandrou:2017dzj,
Alexandrou:2018pbm,Lin:2018pvv,
Alexandrou:2018own,Alexandrou:2019lfo,
Gao:2026wlz}, and transversity~\cite{Alexandrou:2017dzj,
Alexandrou:2018eet,Alexandrou:2018own,
Liu:2018hxv,Alexandrou:2019lfo,
LatticeParton:2022xsd,Gao:2023ktu} case. Many of the early calculations identified the main methodology-related systematic uncertainties that shaped later work.
A substantial part of the early literature on lattice calculations focused on the theoretical and technical infrastructure needed to make these extractions reliable. For example, renormalization prescriptions for Wilson-line operators were developed, including treatments of the Wilson-line power divergence, operator mixing, finite-volume effects, lattice artifacts associated with spatially nonlocal operators, evolution of PDFs, improved interpolators and gauge-fixing precision were also studied~\cite{Monahan:2016bvm,Constantinou:2017sej,Alexandrou:2017huk,Green:2017xeu,Briceno:2017cpo,Ishikawa:2017iym,Briceno:2018lfj,Zhang:2018ggy,Lin:2019ocg,Green:2020xco,Briceno:2021jlb,Constantinou:2022aij,Dutrieux:2023zpy,Zhang:2024omt,Zhang:2025hyo}, along with attempts to benchmark challenges~\cite{Cichy:2019ebf,DelDebbio:2020rgv} for the next generation of calculations.

A second major direction in the late 2010s was the implementation of the SDF approach, which introduced pseudo-distributions and reduced Ioffe-time distributions as coordinate-space quantities from which light-cone PDFs can be extracted~\cite{Orginos:2017kos}. Subsequent lattice calculations applied this framework to nucleon unpolarized PDF and moved toward lighter pion masses and improved control over lattice systematics~\cite{Joo:2019jct,Joo:2020spy}. The finite range and finite precision of lattice matrix elements also motivated dedicated work on reconstruction and moment extraction. Methods based on Ioffe-time moments, Bayesian reconstruction, neural networks, and Bayes-Gauss-Fourier transforms were developed to stabilize the inverse problem and quantify the resolution limitations of lattice data~\cite{Karpie:2018zaz,Karpie:2019eiq,Alexandrou:2020tqq,Medrano:2025cmg}. These studies made clear that the reconstruction of the $x$ dependence suffers from an ill-posed inverse problem whose solution depends on assumptions about smoothness, support, small- and large-$x$ behavior, and the treatment of limited Ioffe-time coverage.

In recent years, lattice calculations of nucleon quark PDFs have progressed toward calculations with better control of systematic uncertainties. In the LaMET framework, an important step was the continuum-limit study of isovector unpolarized and helicity quasi-PDFs at a pion mass of about $370$ MeV~\cite{Alexandrou:2020qtt}. This calculation showed that discretization effects can be significant for nonlocal Wilson-line operators and that continuum extrapolations are essential for quantitatively reliable PDFs. In fact, Ref.~\cite{Green:2020xco} has shown that nonlocal operators suffer from ${\cal O}(a)$ effects. Subsequent work extended LaMET calculations in several directions. Flavor-decomposed calculations included disconnected diagrams and provided separate information on the up, down, and strange contributions to helicity PDFs, and later to unpolarized, helicity, and transversity PDFs~\cite{Alexandrou:2020uyt,Alexandrou:2021oih}. Further progress in LaMET has focused on improved perturbative matching, renormalization, and continuum extrapolations at the physical point. The unpolarized isovector proton PDF has been calculated at physical quark masses with NNLO matching, using both the pseudo-PDF and LaMET frameworks and also extracting low Mellin moments from the same matrix elements~\cite{Gao:2022uhg}. For polarized distributions, the transversity PDF has been determined in the continuum and physical-mass limits using LaMET with hybrid/self-renormalization~\cite{LatticeParton:2022xsd}, while the helicity PDF has been studied in the continuum limit using self-renormalization, renormalization-group resummation, and leading-renormalon resummation~\cite{Holligan:2024wpv}. These calculations show the increasing emphasis on reducing perturbative, renormalization, and lattice-spacing uncertainties, rather than only demonstrating the feasibility of the method.

The SDF program has also developed significantly in recent years. A calculation at physical quark masses provided the first pseudo-PDF determination of the flavor-nonsinglet unpolarized nucleon PDF directly at the physical point~\cite{Bhat:2020ktg}. Other studies used distillation to improve the statistical quality of Ioffe-time pseudo-distributions and the control of excited states~\cite{Egerer:2021ymv}, and performed continuum-limit analyses of unpolarized pseudo-PDFs, including studies of leading-twist separation, discretization effects, two-loop matching, and the dependence on the maximum Wilson-line length~\cite{Karpie:2021pap,Bhat:2022zrw}. The pseudo-PDF approach has also been extended to spin-dependent distributions: the transversity PDF was extracted using NLO matching and Jacobi-polynomial reconstruction~\cite{HadStruc:2021qdf}, and nonsinglet helicity PDFs were determined using distillation, short-distance matching, and a treatment of additional $z^2$ contaminations~\cite{HadStruc:2022nay}. 

The reconstruction of the $x$ dependence has likewise become a major focus. Neural-network methods have been applied to Ioffe-time pseudo-distribution data in an NNPDF-style analysis~\cite{DelDebbio:2020rgv}, and also combined quasi- and pseudo-distributions in a more recent work~\cite{Chu:2025jsi}. Nonparametric approaches use Gaussian-process priors to regularize the inverse problem and provide controlled uncertainty estimates~\cite{Dutrieux:2024rem,Medrano:2025cmg}. These methods are designed to reduce model bias compared with restrictive analytic parametrizations, while making explicit the limited resolution imposed by the finite Ioffe-time range. In parallel, the interface between lattice calculations and phenomenology has been developed through global analyses that incorporate lattice matrix elements together with experimental data~\cite{Bringewatt:2020ixn}, and through window observables designed to benchmark lattice and global-fit PDFs in regions where both methods are reliable~\cite{Karpie:2025ikc}.

In addition to reconstructing the full $x$ dependence, nonlocal matrix elements can be used to extract Mellin moments through their short-distance behavior. This idea was first demonstrated in Ref.~\cite{Karpie:2018zaz}, which showed that moments of PDFs can be obtained from nonlocal Ioffe-time matrix elements, and was applied to continuum-limit results matched in the $\overline{\rm MS}$ scheme. This strategy provides a useful complement to direct PDF reconstruction, because it avoids the ill-posed inverse problem and instead uses the polynomial dependence of short-distance matrix elements on the Ioffe time to isolate moments. It also provides a bridge to the traditional local-operator moment program. For the proton, this nonlocal-OPE strategy has begun to be applied together with $x$-dependent PDF extractions. In Ref.~\cite{Gao:2022uhg}, the lowest Mellin moments of the unpolarized isovector proton PDF were extracted from the leading-twist OPE of nonlocal matrix elements at physical quark masses, in parallel with an NNLO determination of the $x$ dependence using both pseudo-PDF and LaMET analyses. More recently, Ref.~\cite{Gao:2026wlz} applied a similar strategy to the isovector helicity PDF, extracting the lowest helicity moments from OPE ratios of quasi-PDF matrix elements. 

A distinct recent direction is the use of Coulomb-gauge boosted correlations. This approach replaces gauge-invariant Wilson-line operators by gauge-fixed boosted correlators, avoiding the Wilson-line power divergence and potentially improving access to large momenta and long-distance information~\cite{Gao:2023lny}. The first exploratory nucleon calculation in this framework has obtained unpolarized, helicity, and transversity PDFs from boosted Coulomb-gauge correlators, with encouraging behavior for valence-like distributions and larger challenges for the full quark channel~\cite{Gao:2026hix}. 

\subsection{Quark PDFs for the pion and kaon}
\label{subsec:results-pion-kaon}

The pion and kaon PDFs have been increasingly studied as methodologies progressed, and first-principle calculations from lattice QCD are well motivated, as they provide valuable first-principle predictions where experimental data are sparse or non-existing. Their partonic structure is experimentally less constrained than that of the proton, especially for the kaon. They also probe the role of chiral symmetry breaking and flavor-SU(3) symmetry breaking in hadron structure. The first calculations of pion PDFs was dominated by LaMET studies of the valence distribution. Early exploratory work demonstrated the feasibility of calculating boosted pion matrix elements and quasi-PDFs on the lattice~\cite{Zhang:2018rls,Shugert:2018pwi,Zhang:2018nsy}. Follow-up calculations improved the treatment of renormalization and lattice systematics, including studies on fine lattices with nonperturbative RI/MOM renormalization~\cite{Izubuchi:2019lyk} and high-statistics calculations at two fine lattice spacings moving toward the continuum limit~\cite{Gao:2020ito}. In parallel, the SDF approach was applied to the pion through reduced Ioffe-time pseudo-distributions~\cite{Joo:2019bzr}. This provided a complementary coordinate-space extraction of the pion valence distribution.

More recently, pion-PDF calculations have moved toward higher perturbative accuracy, improved systematics, and closer contact with phenomenology. A lattice calculation of the pion valence distribution with NNLO perturbative matching was presented in Ref.~\cite{Gao:2021dbh}, using fine lattice spacings and the hybrid renormalization framework. This was followed by a continuum-extrapolated NNLO calculation at the physical point~\cite{Gao:2022iex}, which combined moment information and $x$-dependent reconstructions and compared the resulting pion PDF with global analyses. Another physical-pion-mass LaMET calculation used high statistics, two-loop perturbative matching, hybrid-ratio and hybrid-RI/MOM renormalization schemes, and resummation improvements to reduce systematic uncertainties~\cite{Holligan:2024umc}. In a separate work, the pion PDF was extracted using the same lattice matrix elements analyzed through both LaMET and SDF~\cite{Miller:2025wgr}.

The kaon PDF has also begun to be studied with nonlocal operators. Ref.~\cite{Lin:2020ssv} presented a LaMET calculation of the kaon valence-quark distribution, together with the corresponding pion valence distribution. Ref.~\cite{Miller:2025wgr} presents the kaon PDF for both LaMET and SDF along the pion. This provides a particularly useful comparison because the two frameworks use closely related nonlocal matrix elements but organize the matching and reconstruction differently. 

Similar to the proton, extracting Mellin moments of pion and kaon PDFs through the short-distance OPE on nonlocal operators has become available. Instead of reconstructing the full $x$ dependence, this approach extracts moments of the PDFs directly from the short-distance behavior of boosted nonlocal matrix elements, providing complementary integral constraints and information on flavor-SU(3) breaking. For the pion, Ref.~\cite{Gao:2022iex} applied the leading-twist expansion to ratios of bilocal pion matrix elements and extracted the second, fourth, and sixth Mellin moments using NNLO Wilson coefficients. This analysis was performed in conjunction with an $x$-dependent reconstruction of the pion valence PDF at the physical point, providing a useful connection between moment information and the reconstructed distribution. More recently, Ref.~\cite{Miller:2026hza} extended the nonlocal-OPE strategy to both the pion and the kaon, using boosted-meson matrix elements of straight-Wilson-line operators within the SDF framework. The analysis studied the dependence on the OPE truncation, the coordinate-space fit window, and the perturbative order of the Wilson coefficients, and used the extracted moments to quantify flavor-$SU(3)$ breaking between the pion and kaon. 

Lattice information on pion PDFs has also begun to enter global QCD analyses. Ref.~\cite{JeffersonLabAngularMomentumJAM:2022aix} incorporated lattice-QCD reduced pseudo-ITDs and current-current correlator matrix elements together with experimental data in a Monte Carlo global analysis of pion PDFs. This type of study is important because it tests how lattice observables can constrain phenomenological PDFs in practice and helps identify which lattice quantities provide the strongest information once their systematic uncertainties are included.

Finally, alternative formulations are being explored. A recent direct momentum-space approach based on Coulomb-gauge quasi-distributions that was first implemented in the proton, was applied to the pion~\cite{Grebe:2026qmt}. This method aims to compute quasi-distributions directly in momentum space and thereby avoid the formal inverse problem associated with reconstructing PDFs from a finite range of coordinate-space data. 

Overall, pion and kaon PDF calculations have progressed from first demonstrations of the pion valence distribution to studies with NNLO matching, continuum and physical-point control, direct kaon calculations, combined LaMET-SDF analyses, nonlocal-OPE moment extractions, and incorporation of lattice data into global analyses. The pion remains the more extensively studied meson, while kaon PDFs are less mature but especially valuable for understanding flavor-SU(3) symmetry breaking in partonic structure.

\subsection{Gluon PDFs}
\label{subsec:results-gluon}

Gluon PDFs play a central role in the momentum and spin structure of hadrons and are essential inputs for collider phenomenology. Their lattice determination is considerably more challenging than that of quark PDFs, because gluonic matrix elements are noisy. Also, the gluon PDF mixes with the quark singlet under perturbative matching and evolution. Despite the challenges, the number of lattice calculations of gluon PDFs has increased significantly in recent years due to major theoretical advances, and includes calculations on the proton, pion, and kaon. In the following, we focus mainly on the proton, where several independent calculations now exist and where both the SDF and LaMET methods have been explored. We also note results for pion~\cite{Fan:2021bcr,Good:2023ecp} and kaon~\cite{Salas-Chavira:2021qku,NieMiera:2025inn} gluon PDFs.

Exploratory works on the gluon PDF for the proton appear in Refs.~\cite{Fan:2018dxu,Fan:2020cpa}. A more recent calculation using the SDF approach was carried out in Ref.~\cite{HadStruc:2021wmh}. This work remains an important reference point because it introduced a complementary strategy for the unpolarized gluon Ioffe-time distribution using distillation, gradient flow, and a summed generalized eigenvalue problem. Distillation allowed the construction of an extended basis of nucleon interpolating operators and improved the treatment of excited states, while gradient flow reduced ultraviolet fluctuations in the gluonic operators. The calculation obtained a statistically controlled Ioffe-time distribution and reconstructed the gluon PDF, which was found to be consistent with phenomenological determinations within uncertainties. The study also emphasized the need for future control of higher-twist effects, finite-volume effects, lattice-spacing effects, unphysical pion-mass effects, and quark-singlet mixing. Using the same setup and computational strategy, the first exploratory lattice calculation of the gluon helicity PDF, $\Delta g(x)$, using SDF was presented in Ref.~\cite{HadStruc:2022yaw}. Although the work did not yet provide a precision determination of $\Delta g(x)$, it demonstrated the feasibility of constructing the matrix elements of the polarized gluon nonlocal operator on the lattice. Follow-up studies explored machine-learning-assisted reconstructions of the gluon helicity distribution and global-analysis frameworks in which lattice data are included together with experimental spin-dependent data~\cite{Khan:2022gluonHelicityML,Karpie:2023nyg}. These studies illustrate two complementary directions: using flexible reconstruction strategies to extract more information from finite lattice data, and incorporating lattice matrix elements directly into global QCD analyses.

A calculation using multiple ensembles implementing the SDF approach is reported in Ref.~\cite{Fan:2022kcb}. This work used a mixed-action setup with clover valence fermions on $N_f=2+1+1$ HISQ ensembles at three lattice spacings ($0.09$, $0.12$, and $0.15$ fm) and different pion masses ($220$ to $700$ MeV). Extrapolations to the continuum limit and physical point were studied along with excited-state effects. The effect of the extrapolation was found to be mild at the available precision, although it increased the final uncertainty. The final distribution was compatible with global analyses in the better-constrained region, while the small-$x$ behavior remained difficult to determine reliably. Quark-gluon mixing was estimated using phenomenological quark PDFs, as lattice estimates for the quark-singlet PDFs have not been calculated.

Another calculation is presented in Ref.~\cite{Delmar:2023agv} using the SDF approach. The frequent use of SDF for gluonic observables is partly motivated by practical considerations. In this framework, one can work with suitable ratios of matrix elements, which help cancel Wilson-line divergences and reduce statistical fluctuations. This is particularly advantageous for gluon operators, for which a direct nonperturbative renormalization based on RI-type vertex functions is substantially more challenging than in the quark case, due to the noisier gluonic operators and the more complicated operator structure. The calculation of Ref.~\cite{Delmar:2023agv} used an $N_f=2+1+1$ maximally twisted-mass ensemble with clover improvement, a pion mass of approximately $260$ MeV, and lattice spacing $0.093$ fm. The analysis explored systematic effects associated with stout smearing of the gluon operator, excited-state contamination, and the dependence on the maximum Wilson-line length entering the fits. A notable new element was the first lattice-based elimination of the mixing with the quark-singlet PDF in the gluon pseudo-PDF formalism. The required quark-singlet nonlocal matrix elements were partly computed on the same ensemble~\cite{Alexandrou:2021oih}, and were extended so they can be analyzed consistently within the pseudo-PDF framework. At the statistical precision of the calculation, the mixing contribution was found to be smaller than the statistical uncertainties, but the work emphasized that such mixing will become increasingly important as gluon-PDF calculations become more precise.
An extension of the work appears in Ref.~\cite{Delmar:2026ltr}, which used the short-distance behavior of the nonlocal gluon matrix elements to extract Mellin moments via an OPE. The work successfully isolated $\langle x^3\rangle_g/\langle x\rangle_g$; higher Mellin moments were compatible with zero. The analysis studied the dependence on the OPE truncation, the range of Wilson-line separations, perturbative scale evolution, and quark-singlet mixing. This calculation illustrates the complementary role of nonlocal operators compared to the local ones: the latter were only able to extract $\langle x\rangle_g$ due to a decaying signal-to-noise ratio, impracticality of renormalization for higher Mellin moments, and mixing with other gluon operators.

Recent advances in renormalization have enabled LaMET studies for gluon PDFs beyond exploratory calculations. A first such attempt is reported in Ref.~\cite{Good:2024lamet}, which used a mixed-action setup with clover valence fermions on an $N_f=2+1+1$ HISQ sea-quark ensemble, with two valence pion masses. This work discusses operator choices, renormalization strategies, pion-mass dependence, and gauge-link smearing effects. The study was completed in Ref.~\cite{Good:2025daz} for the light-cone quantities, which were not included in the first calculation. It used hybrid-renormalized gluon matrix elements and provided a proof of principle of implementing LaMET for gluonic contributions. Refs.~\cite{NieMiera:2025mwj,NieMiera:2025vcx} continue the study of LaMET for the gluon PDF of the proton by applying hybrid renormalization with self-renormalization and gradient-flow smearing on the gluonic operators. Results on different ensembles were analyzed in order to study the continuum limit and pion-mass dependence.

An independent work using LaMET is presented in Ref.~\cite{ChenChen:2025amm} for the unpolarized nucleon gluon PDF with both continuum and infinite-momentum extrapolations. This work used $N_f=2+1$ CLQCD ensembles at three lattice spacings and nucleon momenta up to about $2$ GeV. Distillation was employed to improve the signal of two-point correlation functions, and the analysis used hybrid renormalization, one-loop perturbative matching, and extrapolations to the continuum and infinite-momentum limits. The main new element was the attempt to take both of these limits within a LaMET gluon-PDF calculation.

Machine-learning methods have also begun to enter the gluon-PDF program. These approaches are motivated by the same reconstruction problem discussed in Sec.~\ref{sec:x-reconstruction}: lattice data are available only over a finite range of separations and hadron momenta. In addition to the gluon-helicity application mentioned above, generative machine-learning methods have been applied to both polarized and unpolarized gluon correlation functions, with the goal of extending information from short-distance lattice data and reducing biases associated with restrictive functional forms~\cite{Khan:2022gluonHelicityML,Chowdhury:2024gluonML}. These studies are encouraging, but their results depend on validation of the learned correlations, treatment of systematic uncertainties, and comparison with conventional reconstruction strategies.

To summarize, the above-mentioned studies show that lattice calculations of gluon PDFs have moved from exploratory demonstrations to increasingly systematic determinations. Pseudo-PDF calculations have provided the first $x$-dependent gluon PDFs and have begun to address physical-continuum extrapolations and quark-singlet mixing. LaMET calculations have developed hybrid and self-renormalization strategies and are beginning to explore continuum and infinite-momentum limits. Gluon helicity studies, machine-learning-assisted reconstructions, direct inclusion of lattice Ioffe-time distributions in global analyses, and nonlocal-OPE moments extend the program beyond the unpolarized $x$-dependent PDF. Nevertheless, gluon PDFs remain less mature than nonsinglet quark PDFs.

\subsection{Results beyond leading twist}
\label{subsec:results-twist3}

Most lattice studies of $x$-dependent PDFs have focused on leading-twist distributions, but there is growing interest in extending the same nonlocal-operator methods to distributions beyond leading twist. Twist-3 PDFs provide information that is not contained in the leading-twist functions, for the proton, $f_1$, $g_1$, and $h_1$. They encode quark-mass effects, transverse-spin structure, and genuine quark-gluon correlations. Examples include the scalar distribution $e(x)$, the transverse-spin distribution $g_T(x)$, and the longitudinal-transverse distribution $h_L(x)$. These distributions are relevant for understanding power-suppressed contributions to hard processes and for connecting collinear PDFs with more general descriptions of spin-orbit and quark-gluon dynamics. Often, twist-3 PDFs separate the Wandzura-Wilczek-type contributions from genuine twist-3 terms. The former are determined by leading-twist PDFs through equations-of-motion relations, while the latter arise from quark-gluon-quark correlation functions. These genuine twist-3 contributions are often referred to as dynamical twist-3 or three-parton correlation functions and generally depend on two independent partonic momentum fractions. They therefore contain information beyond what can be inferred from leading-twist PDFs alone. Further information on the implications of twist in hadron structure can be found in this Encyclopedia's chapter ``Higher twist effects in QCD''. 

The first lattice-QCD calculation of an $x$-dependent twist-3 PDF was performed for the isovector distribution $g_T(x)$ of the proton using the quasi-PDF approach~\cite{Bhattacharya:2020cen}. This calculation used $N_f=2+1+1$ twisted-mass fermions with a clover term, a lattice spacing of about $0.093$ fm, a pion mass of about $260$ MeV, and proton boosts up to about $1.67$ GeV. The nonlocal matrix elements were renormalized nonperturbatively in an RI-type scheme and matched to the light-cone distribution in the $\overline{\rm MS}$ scheme at $2$ GeV. The leading-twist helicity distribution $g_1(x)$ was computed in the same setup, allowing a direct comparison of $g_T(x)$ with its Wandzura-Wilczek approximation. The study demonstrated the feasibility of accessing twist-3 PDFs from lattice QCD and found that the Wandzura-Wilczek approximation provides a good description over a broad range of $x$ within the uncertainties of the calculation. A subsequent lattice calculation addressed the chiral-odd twist-3 distribution $h_L(x)$~\cite{Bhattacharya:2021moj}, on the same ensemble. In addition to the isovector combination, the calculation combined the isovector and isoscalar $h_L$ to disentangle the individual up- and down-quark contributions. As in the case of $g_T(x)$, the results provided first evidence that twist-3 PDFs can be studied with nonlocal operators in a Euclidean lattice.

These lattice calculations were accompanied by theoretical developments in the matching and factorization of twist-3 PDFs. The one-loop matching for the chiral-even twist-3 distribution $g_T(x)$ was derived in Ref.~\cite{Bhattacharya:2020xlt}. This work extended the LaMET matching formalism beyond twist two and highlighted the role of singular zero-mode contributions. The matching for the remaining twist-3 PDFs $e(x)$ and $h_L(x)$ was studied in Ref.~\cite{Bhattacharya:2020jfj}, where the treatment of zero-mode contributions was shown to be essential for a consistent relation between quasi-distributions and light-cone twist-3 PDFs. Further theoretical clarification came from factorization analyses of twist-3 axial-vector and chiral-odd nonlocal correlators. Ref.~\cite{Braun:2021aon} derived the factorization formula for the transverse axial-vector quasi-distribution, relevant for $g_T(x)$, in terms of twist-two and twist-3 collinear distributions at one-loop accuracy. The results were formulated in both position and momentum space. Ref.~\cite{Braun:2021gvv} provided the corresponding analysis for chiral-odd quark-antiquark correlations relevant for $h_L(x)$ and $e(x)$, as well as for the leading-twist transversity distribution. An important aspect of this work is that the twist-two component of the $h_L$ quasi- or pseudo-distribution can be separated from the genuine twist-3 part.

In addition to $x$-dependent twist-3 PDFs, moments of twist-3 distributions provide complementary information. A recent calculation extracted the twist-3 moment $d_2$ associated with $g_T(x)$ using an OPE analysis of lattice matrix elements at physical quark masses~\cite{Gao:2026wlz}. This quantity is related to quark-gluon correlations and is often interpreted in connection with the average color Lorentz force acting on the struck quark. The calculation obtained $d_2^{u-d}$ in the $\overline{\rm MS}$ scheme and illustrates how moment information from the OPE can complement direct reconstructions of twist-3 PDFs. Overall, lattice studies of twist-3 PDFs are still at an earlier stage than leading-twist calculations, but they have already established the basic feasibility of the program.

\section{Summary and prospects}
\label{sec:summary-prospects}

The study of PDFs from lattice QCD has entered a qualitatively new stage. For many years, first-principles information on partonic structure was restricted primarily to the lowest Mellin moments obtained from local operators. The development of methods based on spatially nonlocal Euclidean matrix elements has greatly expanded this scope. Large momentum effective theory, short-distance factorization, and related approaches now provide access not only to integral constraints, but also to the $x$ dependence of quark and gluon distributions. This has transformed lattice QCD from a source of a few benchmark quantities into a framework capable of systematically addressing the partonic structure of hadrons in much broader detail. It should be emphasized that extracting PDFs from lattice QCD is not obtained from a single step, but from a chain of controlled theoretical and numerical ingredients, as discussed in this chapter. The calculation requires boosted-hadron matrix elements, renormalization of nonlocal operators, perturbative matching to light-cone quantities, treatment of finite-momentum and finite-distance effects, and reconstruction of continuous distributions from finite Euclidean data. Considerable theoretical progress has been made in each of these areas. 

The current results already demonstrate the range of observables that lattice QCD can now address. Nonsinglet quark PDFs of the proton remain the most mature sector, with calculations now addressing physical quark masses, continuum extrapolations, flavor decomposition, helicity and transversity distributions, and nonlocal-OPE moment extractions. Pion and kaon PDFs provide an especially important arena where lattice QCD can make first-principles predictions in regions where experimental constraints are limited. The comparison between pion and kaon PDFs gives direct access to flavor-$SU(3)$ breaking in partonic structure, while combined LaMET and SDF analyses provide valuable tests of methodology-dependent uncertainties. Gluon PDFs, although still more challenging, have also advanced substantially. Another important direction is the extension of the nonlocal-operator program beyond leading-twist quark PDFs. At the same time, the remaining challenges are clear. Reliable lattice PDFs require simultaneous control over excited-state contamination, finite-volume effects, lattice-spacing artifacts, finite-momentum or finite-distance corrections, perturbative truncation, renormalization-scheme dependence, and the inverse problem in the reconstruction of the $x$ dependence. No single calculation can eliminate all of these uncertainties at once, and comparisons among methods are therefore essential. 

The interface between lattice QCD and global analyses is one of the most promising directions for the coming years. Lattice calculations can provide information on flavor combinations, polarization channels, gluon distributions, and meson PDFs that are difficult to determine from experiment alone. On the other hand, global analyses provide a natural framework for combining lattice observables with experimental data while propagating uncertainties consistently. This connection is particularly important for most of the quantities discussed in this chapter which are not easily accessible experimentally, and for regions of $x$ where experimental data are sparse. As lattice uncertainties become more complete, lattice input can become an increasingly quantitative part of global PDF determinations.

The future of the field of PDFs from lattice QCD is highly encouraging. Lattice-QCD calculations of PDFs have moved beyond proof-of-principle studies and are approaching the level of precision and systematic control needed for phenomenological impact. The field is still developing, and important challenges remain, but the combination of improved algorithms, new computational resources, higher-order perturbative inputs, refined reconstruction methods, and integration within global analyses demonstrate how lattice QCD may play an essential role in mapping the partonic structure of hadrons. 

\section*{Acknowledgements}
We thank the Editors of the Encyclopedia for the invitation to contribute this chapter and for their guidance during its preparation. The rapid progress presented in this chapter reflects the efforts of a broad and active community, and we thank its members for many stimulating discussions over the years that inspired this review.

M.~C. acknowledges financial support from the U.S. Department of Energy, Office of Nuclear Physics,  under Grant No.\ DE-SC0025218. K.~C. is supported by the National Science Centre (Poland) grant OPUS No. 2021/43/B/ST2/00497.

\bibliographystyle{Harvard}
\bibliography{bibliography}

\end{document}